\documentstyle[11pt,paspconf]{article}
\begin{document}
\hoffset -1.0truein
\voffset -0.5truein
\rightline{To Appear in {\sl Energy Transport in Radio Galaxies
and Quasars}}
\rightline{eds. P. Hardee, A. Bridle, \& A. Zensus (ASP Conference
Series) 1996}
\vskip 0.143truein

\title{Estimates of Doppler Factors, Outflow Angles, and Bulk Lorentz
Factors for a Sample of Compact Radio Sources}
\author{R. A. Daly, E. J. Guerra, \& A. Guijosa}

\affil{Department of Physics, Princeton University, Princeton, NJ
08544}

\begin{abstract}
Compact radio sources can be strongly affected by relativistic
motion.  The relativistic motion is characterized by 2
quantities: the speed or bulk Lorentz factor of the outflow, and
the angle subtended by the outflow direction and the line of
sight to the observer.  One combination of these yields
the Doppler factor of the outflow, and a different combination
yields the apparent (sometimes superluminal)
speed with which a feature moves away from
the core.  Since there are two observed quantities, the Doppler
factor and the apparent velocity, the two unknowns, the bulk
Lorentz factor and the outflow angle, can be estimated
separately.  Thus, the bulk Lorentz factor and the outflow
angle can be estimated separately for a source that has an estimate
of both the Doppler factor and an apparent velocity, assuming
that the apparent velocity is due to physical motion rather than a
pattern speed.

Two key results are presented in this paper. First, equipartition
Doppler factors have been estimated for a sample of 105 sources
with previously estimated inverse Compton Doppler factors.  The
different categories of active galactic nuclei (AGN) have
different ranges and typical values of the Doppler factor; lobe
dominated AGN tend to have low Doppler factors, core-dominated quasars
tend to have large Doppler factors, and BL Lacertae objects span
the full range of Doppler factors.  These results support the
``orientation unified model" for different categories of AGN.

The second key result comes from combining equipartition
Doppler factors and apparent velocities from superluminal motion
studies.  There are 43 sources for which apparent speeds and
equipartition Doppler factors are available.  The bulk Lorentz factor and
outflow angle are estimated for each of these 43 sources.
Again, typical
parameter values for different categories of AGN are quite
interesting, and generally agree with expectations based on the
orientation unified model.

\end{abstract}


\section{Introduction}

Relativistic motion in compact radio sources is indicated by apparent
superluminal motion, one-sidedness of jets, rapid flux
variations, and a deficit of synchrotron self-Compton X-rays.
The relativistic motion is characterized by 2 quantities:
the angle
$\theta$ subtended by the direction of outflow and the line of
sight to the observer, and the
bulk Lorentz factor of the outflow $\gamma$, where
$\gamma=1/\sqrt{1-\beta^2}$, and
$\beta$ is the speed of the outflow relative to the speed
of light.

The Doppler factor $\delta$ of the outflow is a combination of these 2
parameters: $\delta = \gamma^{-1} (1-\beta \cos{\theta})^{-1}$.
The apparent speed $\beta_{app}$ of a particular feature depends upon a
different combination of these 2 parameters:
$\beta_{app} = \beta \sin{\theta} (1-\beta \cos{\theta})^{-1}$.
If a source has an estimate of both $\delta$ and $\beta_{app}$,
then there are 2 observations to constrain the 2 unknowns, so
the bulk Lorentz factor $\gamma$ and the outflow angle $\theta$
may be estimated separately for the source.  Combining the
equations for $\beta_{app}$ and $\delta$, we obtain:

\begin{equation}
\gamma=(2 \delta)^{-1}(\beta_{app}^2+ \delta^2 + 1)
\end{equation}

and
\begin{equation}
\cos{\theta}=\left({1 \over \sqrt{\gamma^2
-1}}\right)(\gamma-\delta^{-1})~.
\end{equation}

Equipartition Doppler factors for 105 sources are discussed in \S 2, bulk
Lorentz factors and outflow angles for 43 sources are discussed
in \S 3.  The results are summarized in \S 4.

\section{Equipartition Doppler Factors}

Bulk relativistic motion can strongly alter the appearance of the
emitter. In order to be able to study intrinsic source
properties, and selection effects that determine which subset of the parent
population make it into a radio catalogue, it is necessary to
understand and correct for relativistic effects.

The Doppler factor may be estimated from single epoch radio
observations using the ``equipartition Doppler factor,"
first introduced by Readhead (1994), and computed for a sample of
105 radio sources by Guijosa and Daly (1996) using
the sample compiled and studied by Ghisellini et al. (1993).
Guijosa and Daly
(1996) compare the equipartition Doppler factor with the inverse
Compton Doppler factor, and find that the two are strongly
correlated.
The appropriate error bar for each of these quantities is
presently under investigation (Guerra and Daly 1995), and we plan
to investigate in detail whether the fact that the ratio of
equipartition to inverse Compton Doppler factor is about
$1 \pm 0.1$ indicates that both are reliable estimators of the
true Doppler factor.

It is quite interesting to note that the range and typical value
of the Doppler factor are different for each of the categories of
AGN studied.  The 8 radio galaxies in the sample all have
equipartition Doppler factors $\delta_{eq}$ less than or on the
order of 1, as do most of the 11 lobe-dominated quasars (except
for one quasar with $\delta_{eq} \simeq 65$).  The 24 core-dominated
high-polarization and 29 core-dominated
low-polarization quasars are quite similar and have
$\delta_{eq}$ that range from about 1 to 30; the only
apparent difference between the populations is that the
core-dominated low-polarization quasars seem to have a cluster of
points with $\delta_{eq} < 1$ that is not present in the
high-polarization population.  And, the 33 BL Lacertae objects have
$\delta_{eq}$ that span the full range, with $\delta_{eq}$ much less
than one up to about relatively large values of about 30
(see figure 1 and the figures of Guijosa and Daly 1996).

\begin{figure}[htb]
\caption{Equipartition Doppler Factor as a Function of (1+z)}
\end{figure}

The range of $\delta_{eq}$ for each category of source is evident
on figure 1, which shows the redshift behavior of $\delta_{eq}$
for the sample of 105 sources discussed by Guijosa and Daly
(1996).  The Doppler factor appears to increase with redshift,
as expected from selection effects; more distant sources must be
more strongly Doppler boosted to make it into the sample.
Note, however, that a
large part of the effect is due to low-redshift BL Lacs,
and the redshift behavior of $\delta_{eq}$ may be different for
different categories of AGN; this is
presently under investigation.

\section{Bulk Lorentz Factors and Outflow Angles}

It would be fascinating and enlightening to be able to study the
bulk Lorentz factor and outflow angle for different categories of
AGN.  This would shed light on the relationships between
different categories of AGN, and allow an estimate of
selection effects for different types of AGN as a function of
redshift.  As mentioned in \S 1, it is possible to estimate
$\gamma$ and $\theta$ separately for sources with an estimate of both
$\beta_{app}$ and
$\delta_{eq}$.
Two important assumptions are: (1) that the apparent velocity is
due to a physical motion of the emitter rather than a pattern
speed of the flow, and (2) that the equipartition Doppler factor
provides a good rough estimate of the true Doppler factor.  Both
of these assumptions are presently under investigation by Guerra and
Daly (1995), who discuss the uncertainty on estimates of
$\delta_{eq}$, $\gamma$, and $\theta$.  Although error bars are
not included here, some obvious trends are indicated by the data
that are likely to stand the test of time.

\begin{figure}[htb]
\caption{Outflow Angle $\theta$ as a Function of Bulk Lorentz
Factor $\gamma$}
\end{figure}

Figure 2 shows the values of $\theta$ and $\gamma$ obtained by
applying equations (1) and (2) to the 43 sources studied by both
Vermeulen and Cohen (1994), who investigate $\beta_{app}$ for
a large number of sources, and
by Guijosa and Daly (1996), who investigate $\delta_{eq}$ for
a large number of sources.
Radio galaxies have Lorentz factors that
are typically close to unity, and tend to have
very large angles to
the line of sight.  Specifically,
6 of the 7 radio galaxies have values of $\gamma$
between about 1 and 2, and the seventh has $\gamma \sim 10$; all
of the sources have $\theta$ between $45^o$ and $180^o$, and
about half of the radio galaxies in the sample
have $\theta$ greater
than $90^o$.  If these values
are accurate, they indicate that these radio galaxies have
a relatively slow, one-sided jet that is
pointing {\it away} from the observer about half the time.

Six of the 7
lobe-dominated quasars in the sample have
$\theta$ between $20^o$ and $60^o$, and one has $\theta \sim
0^o$; the Lorentz factors range from 2 to 30, with an apparent
peak at $\gamma \sim 10$.

Nine of the 10 low-polarization core-dominated quasars have
$\theta$ between $0^o$ and $20^o$, and one has $\theta \sim
45^o$; the $\gamma$ factors range from 2 to 20 with an apparent
peak at $\gamma \sim 10$.  The range of Lorentz factors for the
high-polarization core-dominated quasars seems to be larger
than that for the low-polarization quasars, with $\gamma$ ranging
from about 1 to 200 (it is not clear if this apparent difference
is significant), and 4 of the 8 high-polarization core-dominated
quasars have $\gamma \sim
10$.  All of the 8 high-polarization core-dominated
quasars in the sample have $\theta$ between $0^o$ and $20^o$.
The 3
core-dominated quasars without polarization information have
values of $\theta$ between about $30^o$ and $60^o$, and
$\gamma$ between about 1 and 10.

The 8 BL Lacertae ojects seem to have a relatively
broad range of $\theta$ and
$\gamma$, and sources with large $\theta$ tend to have small
$\gamma$.  Two BL Lacs have $\theta$ between $40^o$ and
$60^o$ and $\gamma$ between 1 and 2,
and 6 BL Lacs have $\theta$ between about zero
and $20^o$ and $\gamma$ between about 2 and 30.

\begin{figure}[htb]
\caption{Outflow Angle $\theta$ as a Function of (1+z)}
\end{figure}

\begin{figure}[htb]
\caption{Bulk Lorentz Factor as a Function of (1+z)}
\end{figure}

Figures 3 and 4 show $\theta$ and $\gamma$ as a function of
$(1+z)$.  Both appear to exhibit the type of selection effect
seen in figure 1 and
expected if the intrinsic luminosity of the sources is
nonevolving, since in this case higher redshift sources must be
more strongly Doppler boosted to make it into the sample.  Thus,
figures 3 and 4 suggest that $\theta$ decreases and $\gamma$
increases with redshift. This is not a result of the adopted
cosmology; similar results are obtained when
different cosmological parameters are assumed.  However, it
is not clear if the effect is significant, or if it
is operating for each category of AGN
separately;
this point is presently under investigation.
Generally speaking, the results presented here are
consistent with those presented by
Vermeulen and Cohen (1994).

\section{Summary}

It is striking that the different categories of AGN have
different typical values and ranges of $\delta_{eq}$, $\gamma$,
and $\theta$.  Generally, different categories of AGN
have values of these
parameters that are consistent with expectations based on
the orientation unified model.  It is quite interesting to
note that radio galaxies, lobe-dominated
quasars, and some BL Lacs
have very similar values of the equipartition Doppler
factor $\delta_{eq}$, but clearly separate out on the
$\theta$-$\gamma$ diagram.  The radio galaxies studied here
appear to lie in a region of the diagram
that is separate from the
other types of AGN.  Further,
the radio galaxies have relatively slow outflow
velocities, and about half of the sources are pointing
{\it away} from the observer, suggesting one-sided
slow jets; note
that radio galaxies are the only category of
source with $\theta > 90^o$.
Lobe-dominated quasars, most BL Lacs, and
core-dominated high- and low-polarization quasars fill in the
rest of the $\theta$-$\gamma$ diagram.

Lobe-dominated quasars have relatively large
$\theta$ and small $\gamma$ compared with other types of quasars,
but lie in a region of the diagram distinct from radio
galaxies.  Core-dominated quasars tend to
have the largest values of $\gamma$ and smallest values
of $\theta$.  BL Lacertae objects appear to span
the full range of parameters between lobe-dominated quasars and
core-dominated quasars on the $\theta$-$\gamma$ diagram
(of course, it is always possible that the category into which an
individual object has been placed may be incorrect).

Similar conclusions are indicated by $\delta_{eq}$.
It is clear from figure 1 that radio
galaxies and lobe-dominated quasars tend to have low values of
the equipartition Doppler factor, typically
$\delta_{eq} < 1$.  Core-dominated high- and low-polarization
quasars tend to have high values of the Doppler factor,
with $\delta_{eq}$ between about 1 and 30, with a few sources
having a value less than 1.  And, the BL Lacertae objects
seem to span the full range of $\delta_{eq}$ from values much
less than one to values over 10.  Again, we see that BL Lacs
span the full range, with values similar to
lobe-dominated quasars and
core-dominated quasars, though now with a much larger sample.

The large range of $\theta$ and $\delta_{eq}$ seen for BL Lacs
may suggest either that we are looking so close to the central engine
that the full collimation is not complete, so we see a ``spray"
and thus a broad range of $\theta$ and $\delta_{eq}$, or that the opening
angle that allows an observer to see into the heart of the AGN
and classify it as a BL Lac
is larger than previously thought,
or perhaps some
sources have been mistakenly classified as BL Lacs.

It is important to keep in mind that the $\theta$-$\gamma$
diagram
has been constructed assuming that $\delta_{eq}$ is a good
measure of the true Doppler factor, and that $\beta_{app}$
reflects a bulk flow rather than a pattern speed. Both of
these assumptions are presently under investigation
(Guerra \& Daly 1995).

\acknowledgments

It is a pleasure to thank the participants of this
meeting and other colleaques
for interesting discussions, especially Marshall Cohen, Ed Groth,
Alan Marscher, Tim Pearson,
Tony Readhead, Larry Rudnick,
Peter Scheuer, Rene Vermeulen, and Paul Wiita.
This work was supported in part by the US National Science
Foundation through a Graduate Student Fellowship and a National
Young Investigator Award,
by the Independent College Fund of New Jersey, and by a grant
from W. M. Wheeler III.

\end{document}